\documentclass{article}
\usepackage{spconf}

\usepackage{amsmath, bbm}
\usepackage{amssymb}
\usepackage{amsfonts}
\usepackage{amsthm}
\usepackage{algorithm}
\usepackage{algorithmic}
\usepackage{nicefrac}
\usepackage{bm}
\usepackage{hyperref}
\usepackage{cite}
\usepackage[utf8]{inputenc} 
\usepackage[T1]{fontenc}
\usepackage{graphicx}
\usepackage[font=small]{caption}
\usepackage{tikz}
\usetikzlibrary{positioning}
\usepackage{pgfplots}
\usepackage{epstopdf}
\usepackage[caption=false,font=footnotesize,labelformat=simple]{subfig}
\usepackage{multirow}
\usepackage{lipsum}
\usepackage{mathtools}

\long\def\comment#1{}

\input{mysymbol.sty}

\title{Delay-aware Backpressure Routing Using Graph Neural Networks}

\name{Zhongyuan Zhao$^\star$, Bojan Radojicic$^\dag$, Gunjan Verma$^\ddag$, Ananthram Swami$^\ddag$, and Santiago Segarra$^\star$
\thanks{Research was sponsored by the Army Research Office and was accomplished under Cooperative Agreement Number W911NF-19-2-0269. 
The views and conclusions contained in this document are those of the authors and should not be interpreted as representing the official policies, either expressed or implied, of the Army Research Office or the U.S. Government. 
The U.S. Government is authorized to reproduce and distribute reprints for Government purposes notwithstanding any copyright notation herein.
\newline
Emails: $^\star$\{zhongyuan.zhao, segarra\}@rice.edu, $^\dag$bojanradojicic00@gmail-.com, $^\ddag$\{gunjan.verma.civ, ananthram.swami.civ\}@army.mil}}
\address{$^\star$Rice University, USA \hspace{10mm} $^\dag$University of Novi Sad, Serbia \\ \hspace{2mm}  $^\ddag$US Army’s DEVCOM Army Research Laboratory, USA}

\begin{document}
\ninept
\renewcommand{\baselinestretch}{0.98}
\maketitle
\begin{abstract}
We propose a throughput-optimal biased backpressure (BP) algorithm for routing, where the bias is learned through a graph neural network that seeks to minimize end-to-end delay.
Classical BP routing provides a simple yet powerful distributed solution for resource allocation in wireless multi-hop networks but has poor delay performance. 
A low-cost approach to improve this delay performance is to favor shorter paths by incorporating pre-defined biases in the BP computation, such as a bias based on the shortest path (hop) distance to the destination. 
In this work, we improve upon the widely-used metric of hop distance (and its variants) for the shortest path bias by introducing a bias based on the link duty cycle, which we predict using a graph convolutional neural network.
Numerical results show that our approach can improve the delay performance compared to classical BP and existing BP alternatives based on \emph{pre-defined} bias while being adaptive to interference density.
In terms of complexity, our distributed implementation only introduces a one-time overhead (linear in the number of devices in the network) compared to classical BP, and a constant overhead  compared to the lowest-complexity existing bias-based BP algorithms.
\end{abstract}
\section{Introduction}\label{sec:intro}

Wireless multi-hop networks have been traditionally used in military communications, disaster relief, and wireless sensor networks, and are envisioned to support emerging applications such as connected vehicles, drone/robot swarms, xG (device-to-device, wireless backhaul, and non-terrestrial coverage), Internet of Things, and machine-to-machine communications \cite{Lin06,sarkar2013ad,kott2016internet,akyildiz20206g,cisco2020,chen2021massive}.
An attractive feature of wireless multi-hop networks is its self-organizing capability without relying on infrastructure, enabled by distributed resource allocation schemes.
Among those schemes, backpressure (BP) routing~\cite{tassiulas1992} is a well-established solution for resource allocation across the physical, media access control (MAC), and network layers \cite{neely2005dynamic,georgiadis2006resource,jiao2015virtual,cui2016enhancing,gao2018bias,Alresaini2016bp,ji2012delay,athanasopoulou2012back,hai2017delay,Rai2017loop,yin2017improving,ying2010combining,Ying2012scheduling,Ryu2012timescale}.
In the BP algorithm, each node maintains a separate queue for packets to each destination (also denominated as commodity), routing decisions are made by selecting the commodity with the maximum differential queue backlog between the two ends of each link, and data transmissions are activated on a set of non-interfering links via MaxWeight scheduling~\cite{tassiulas1992}.
The BP mechanism can drive the packets to explore all possible routes towards their destinations, while stabilizing the queues in the network for any flow rates within the network capacity region, i.e., BP achieves throughput optimality \cite{tassiulas1992,neely2005dynamic,georgiadis2006resource}. 

\begin{figure}
    \centering
    \vspace{-0.2in}
    \hspace{-5mm}
    \subfloat[]{
    \includegraphics[height=1.3in]{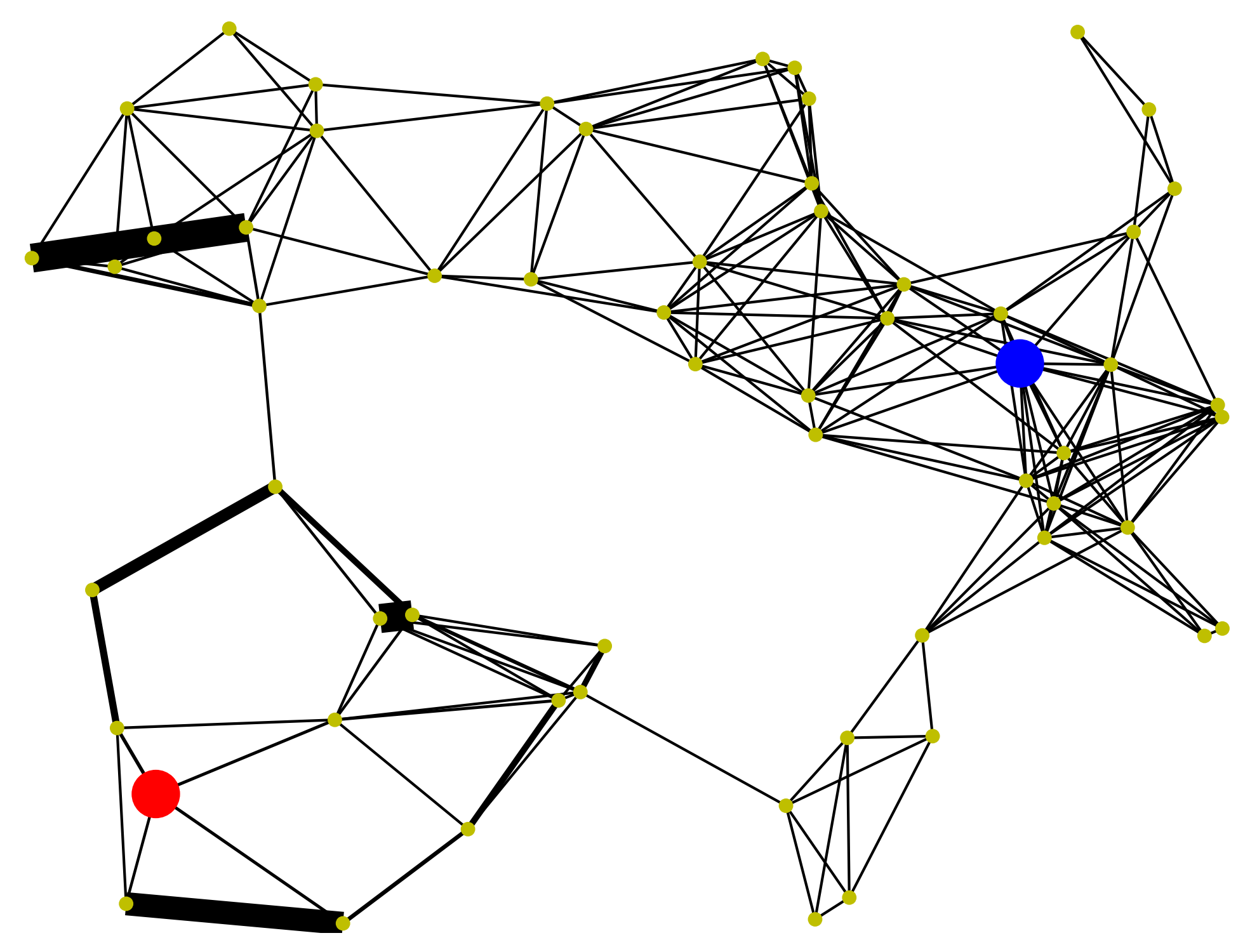}
    \label{fig:motivation:basic}\vspace{-0.05in}
    } \hspace{-3mm}
    \subfloat[]{
    \includegraphics[height=1.3in]{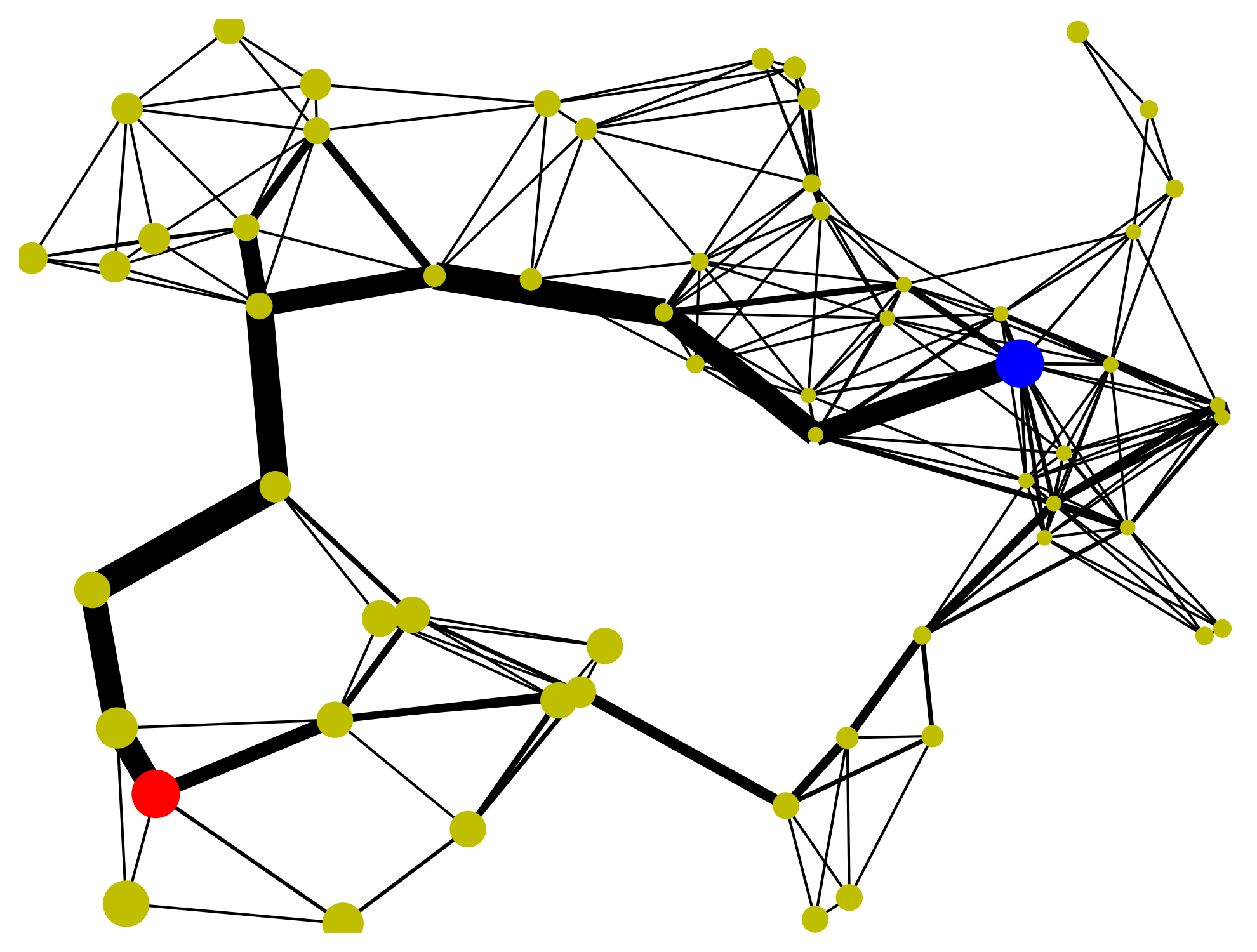}
    \label{fig:motivation:hop}\vspace{-0.05in}
    }
    \vspace{-0.1in}
    \caption{{\small A flow from the red node to the blue node in a wireless multi-hop network with 60 nodes. 
    The normalized (across all links) number of packets sent over each link in 500 time slots is illustrated by their width. 
    (a) Basic BP routing. (b) Enhanced dynamic BP routing (EDR) \cite{neely2005dynamic,georgiadis2006resource} with a pre-defined bias (visualized by the node sizes) given by a scaled version of the shortest path to the destination.}
    }
    \label{fig:motivation}
    \vspace{-0.2in}
\end{figure}

However, it is well-known that the classical BP routing suffers from poor delay performance for flows of low to medium rate  \cite{neely2005dynamic,georgiadis2006resource,jiao2015virtual,cui2016enhancing,gao2018bias}, exhibiting undesirable characteristics such as \emph{slow startup}, \emph{random walk}, and the \emph{last packet problem} \cite{Alresaini2016bp,ji2012delay}. 
When a flow starts, many packets have to be first backlogged to form stable queue backlog-based gradients, causing large initial end-to-end delay.
During BP scheduling, the fluctuations in queue backlogs drive packets towards random directions, causing unnecessarily long routes or loops.
The phenomena of slow startup and random walk in BP routing are illustrated by the example in Fig.~\ref{fig:motivation:basic}, in which packets from the red source node did not reach their blue destination in the first 500 time slots, but were trapped in two loops shown by the thickest edges.

The existing efforts to improve the delay performance of BP can be categorized into four types: 
1) Use pre-defined queue-agnostic biases, such as the shortest path distance \cite{neely2005dynamic,georgiadis2006resource} or functions of it \cite{jiao2015virtual}, to guide low-rate flows through the shortest routes.
For example, with BP enhanced by the shortest path distance bias \cite{neely2005dynamic,georgiadis2006resource}, in Fig.~\ref{fig:motivation:hop} packets can quickly reach their destination through two major routes around the empty central area.
2) Use queue-dependent biases that aggregate the queueing state information (QSI) of the local neighborhood (or global QSI) to improve the myopic BP decisions \cite{cui2016enhancing,gao2018bias} or use shadow queues \cite{athanasopoulou2012back,Alresaini2016bp} to dynamically increase the backpressure.
3) Use delay metrics to replace the queue-dependent biases \cite{ji2012delay,cui2016enhancing,hai2017delay} or dynamically select routing schemes~\cite{yin2017improving}.
4) Impose restrictions on the routes~\cite{Rai2017loop} or hop counts~\cite{ying2010combining} to reduce the effect of loops.
Solutions based on queue-agnostic biases are relatively simple, effective in mitigating the slow startup and random walk problems, they are throughput optimal~\cite{neely2005dynamic,georgiadis2006resource,jiao2015virtual}, and cost only a one-time communication overhead.
In contrast, leveraging neighboring QSI incurs additional overhead per time slot, using delay metrics \cite{ji2012delay,cui2016enhancing,hai2017delay} and methods based on route restrictions \cite{Rai2017loop,ying2010combining} are more complex in implementation and could reduce the network capacity region.
In addition, most of the aforementioned approaches require careful parameter tuning, typically done via trial-and-error.
Besides those efforts, practical concerns in network connectivity \cite{Ryu2012timescale,Alresaini2016bp} and network state uncertainty \cite{Ying2012scheduling} have also been addressed.

In this work, our goal is to maintain the simplicity and computational tractability of the first type of existing works (queue-agnostic biases) but improve their performance through graph-based machine learning.
Specifically, we seek to improve the widely-used bias given by the shortest hop distance to the destination \cite{neely2005dynamic,georgiadis2006resource,jiao2015virtual,yin2017improving,ying2010combining}, by considering the differences in link duty cycles in scheduling, rather than treating all links equally. 
We propose to predict the scheduling duty cycles of all the links with a graph neural network (GNN), an architecture that has been recently applied to improve the throughput \cite{zhao2021icassp,zhao2022j}, delay \cite{zhao2022icassp_a}, and overhead \cite{zhao2022icassp_b} of link scheduling in wireless multi-hop networks.
From the perspective of routing, the interfering wireless links are transformed into conflict-free links with effective rates equal to the link rates weighted by scheduling duty cycles, which can improve the routing decision, e.g., by avoiding hot-spots with many interfering neighbors (high betweenness centrality \cite{Katsaros2010social}).
In addition, our approach can also lift the burden of parameter tuning by automatically increasing the weight of pre-defined biases in wireless networks with high interference degree.


\vspace{1mm}
\noindent
{\bf Contribution.} The contributions of this paper are twofold:\\
1) We propose an approach to leverage the scheduling duty cycle of links for delay-aware
BP routing by learning appropriate biases using GNNs, 
and \\
2) Through numerical experiments, we demonstrate the superior delay performance of the shortest path bias based on link duty cycle in BP routing, especially in wireless networks with high interference.

\section{System Model}
\label{sec:problem}

A wireless multi-hop network can be modeled as an undirected graph $\ccalG^{n}=(\ccalV, \ccalE)$, where $\ccalV$ is a set of nodes representing user devices in the network, and $\ccalE$ represents a set of links, where $e=(i,j)\in\ccalE$ for $i,j\in\ccalV$ represents that node $i$ and node $j$ can talk to each other.
We call $\ccalG^{n}$ the connectivity graph, as it describes the connectivity relationship of the network.
Here, we assume that $\ccalG^{n}$ is a connected graph, so that two arbitrary nodes in the network can always reach each other.
Notice that routing involves directed links, so we use ($\overrightarrow{i,j}$) to denote data packets being transmitted from node $i$ to node $j$ over link $(i,j)$.
There is a set of flows $\ccalF$ in the network, in which a flow $f=(i,c)\in\ccalF$, where $i\neq c$ and $i,c\in\ccalV$, describes the stream of packets from a source node $i$ to a destination node $c$, potentially through multiple links.
At a node $i\in\ccalV$, there is a set of queues, $\{U_{i}^{(c)}|c\in\ccalV\}$, in which $U_{i}^{(c)}$ represents the length of the queue for data packets destined to node $c$ (or packets of commodity $c$).

To describe the conflict relationship between links, we define the \emph{conflict graph}, $\ccalG^c=(\ccalE,\ccalC)$, as follows: a vertex $e\in\ccalE$ represents a link in the network, and the presence of an undirected edge $(e_1, e_2)\in\ccalC$ captures the interference relationship between links $e_1, e_2\in\ccalE$.
We will be focusing on two popular models often used to define the conflict relationship between two links: 
1)~Interface conflict, where the conflict graph is given by the line graph of the connectivity graph.
This represents the case where two links sharing the same node cannot be turned on simultaneously, e.g., if each node is equipped with only one radio transceiver.
2)~Physical distance interference model~\cite{cheng2009complexity}, which arises when two links interfere with each other if their incident users are within a certain distance such that their simultaneous transmission will cause the outage probability to exceed a prescribed level.
A simplified scenario where all the users transmit at identical power levels with an omnidirectional antenna can be captured by the unit-disk interference model. 
In this model, two links conflict with each other if any of their nodes are within a pre-defined distance, which is the same for every pair of links.
For the rest of this paper, we assume the conflict graph $\mathcal{G}^{c}$ to be known, e.g., by each link monitoring the wireless channel, or through more sophisticated estimation as in~\cite{yang2016learning}. 

The MAC of the wireless network is assumed to be time-slotted orthogonal multiple access. 
Each time slot $t$ contains a stage of decision making for routing and scheduling, followed by the second stage of data transmission.
Therefore, we use $U_{i}^{(c)}(t)$ to describe the queue of commodity $c$ at node $i$ at the beginning of time slot $t$.
The exogenous packet arrivals are collected by the non-negative integer matrix $\bbA\in\mathbb{Z}_{+}^{|\ccalF|\times T}$, in which the element of row $f$ and column $t$, $\bbA_{f,t}$, is the number of packets arriving at the source node of flow $f$ at time slot $t$.
Matrix $\bbR\in\mathbb{Z}_{+}^{|\ccalE|\times T}$ collects the (stochastic) real-time link rates, of which an element $\bbR_{e,t}$ represents the number of packets that can be delivered over link $e$ in time  slot $t$.
The long term link rate of a link $e\in\ccalE$ is denoted by  $r_{e}=\mathbb{E}_{t\leq T}\left[\bbR_{e,t}\right]$, 
and $\bar{r}=\mathbb{E}_{e\in\ccalE,t\leq T}\left[\bbR_{e,t}\right]$ is the network-wide average link rate.

\section{(Biased) Backpressure Routing}
\label{sec:basic}

Classical backpressure routing is a distributed algorithm for routing and scheduling, consisting of 4 steps.
First, for each {directed} link ($\overrightarrow{i,j}$), the BP algorithm selects the optimal commodity $c_{ij}^{*}(t)$ as the one with the maximal backpressure, which is defined as the difference of queue lengths between the sender $i$ and the receiver $j$,
\begin{equation}\label{E:commodity}
    c_{ij}^{*}(t) =\argmax_{c\in\ccalV}\{ U_{i}^{(c)}(t) - U_{j}^{(c)}(t) \} \;.
\end{equation}
In step 2, the maximum differential backlog of ($\overrightarrow{i,j}$) is found as:
\begin{equation}\label{E:weight}
    w_{ij}(t) =\max\{ U_{i}^{(c_{ij}^*(t))}(t) - U_{j}^{(c_{ij}^*(t))}(t), 0 \}\;.
\end{equation}
In step 3, MaxWeight scheduling \cite{tassiulas1992} finds the schedule $\bbs (t)\in\{0,1\}^{|\ccalE|}$ to activate a set of \emph{non-conflicting links} achieving the maximum total utility, where the per-link utility is  $u_{ij}(t)=\bbR_{ij,t}\tilde{w}_{ij}(t)$, 
\begin{equation}\label{E:scheduling}
    \bbs (t) = \argmax_{\tilde{\bbs} (t)\in\{0,1\}^{|\ccalE|}} \tilde{\bbs}(t)^\top  \left[\bbR_{*,t}\odot\tilde{\bbw}(t)\right] \;,
\end{equation}
where vector $\bbR_{*,t}$ collects the real-time link rate of all links, vector $\tilde{\bbw}(t)=\left[ \tilde{w}_{ij}(t) | (i,j)\in\ccalE\right]$, where $\tilde{w}_{ij}=\max\{w_{ij}(t),w_{ji}(t)\}$, and the direction of the link selected by the $\max$ function will be recorded for step 4.
Notice that the MaxWeight scheduling in \eqref{E:scheduling} involves solving a maximum weighted independent set (MWIS) problem on the conflict graph (due to the implicit non-conflict constraint on the schedule), which is NP-hard \cite{joo2010complexity}. 
Therefore, in practice, \eqref{E:scheduling} is solved approximately by heuristics, such as centralized greedy maximal scheduler (GMS), distributed local greedy scheduler (LGS)~\cite{joo2012local}, and GNN-enhanced schedulers~\cite{zhao2022j}.
In step 4,  all of the real-time link rate $\bbR_{ij,t}$ of a scheduled link is allocated to its optimal commodity $c_{ij}^{*}(t)$.
The final transmission and routing variables of commodity $c\in\ccalV$ on link ($\overrightarrow{i,j}$) is
\begin{equation}\label{E:quota}
    \mu_{ij}^{(c)}(t) = \begin{cases}
         \bbR_{ij,t}, & \text{if } c=c_{ij}^{*}(t), w_{ij}(t)>0, s_{ij}(t)=1, \\
         0, & \text{otherwise}.
    \end{cases}
\end{equation}
When a set of pre-defined biases for each pair of node and commodity, $\ccalB=\{B_{i}^{(c)}|i,c\in\ccalV\}$, are used to improve the delay performance \cite{neely2005dynamic,georgiadis2006resource,jiao2015virtual}, step 1 in \eqref{E:commodity} and step 2 in \eqref{E:weight} now respectively become,
\begin{equation}\label{E:commodity:bias}
    c_{ij}^{*}(t) =\argmax_{c\in\ccalV}\{ \tilde{U}_{i}^{(c)}(t) - \tilde{U}_{j}^{(c)}(t) \} \;,
\end{equation}
\begin{equation}\label{E:weight:bias}
    w_{ij}(t) =\max\{ \tilde{U}_{i}^{(c_{ij}^*(t))}(t)  - \tilde{U}_{j}^{(c_{ij}^*(t))}(t), 0 \}\;,
\end{equation}
where $\tilde{U}_{i}^{(c)}(t) = U_{i}^{(c)}(t) + B_{i}^{(c)}$. 
Since the bias $B_{i}^{(c)}$ is considered to be independent from the queue lengths, we can see it as a non-negative \emph{constant}, i.e., it does not have to be updated in every time slot.
In practice, $\ccalB$ can still be updated from time to time to match the changing network topology.

\section{Link Duty Cycle Prediction with GNNs}
\label{sec:solution}

In MaxWeight scheduling, the likelihood of a link being scheduled depends on its local conflict or interference topology and the network traffic load.
We propose to use a GNN to predict the expected delay on each link, which can serve to design a better bias in~\eqref{E:commodity:bias}-\eqref{E:weight:bias} than simply the shortest path distance (in number of hops) between node $i$ and node $j$ for commodity $c$.

We propose to predict the link duty cycle $\bbx\in\reals^{|\ccalE|}$,  as $\bbx = \Psi_{\ccalG^c}(\boldsymbol{1};\mathbf{\bbomega})$, where  $\Psi_{\ccalG^c}$ is an $L$-layered convolutional GNN defined on the conflict graph $\ccalG^c$, and $\bbomega$ is the collection of trainable parameters of the GNN.
We define the output of an intermediate $l$-th layer of the convolutional GNN as $\bbX^l \in\reals^{|\ccalE|\times g_{l}}$, $\bbX^0 = \boldsymbol{1}^{|\ccalE|\times 1}$, $\bbx = \bbX^L_{*0} $ (column $0$ of $\bbX^L$), and 
the $l$-th layer of the GNN is expressed as
\begin{equation}\label{E:gcn}
	\mathbf{X}^{l} = \sigma_l\left(\mathbf{X}^{l-1}{\bbTheta}_{0}^{l}+\bbcalL \mathbf{X}^{l-1}{\bbTheta}_{1}^{l}\right), \; l\in\{1,\dots,L\}.
\end{equation}
In~\eqref{E:gcn},  
$\bbcalL$ is the normalized Laplacian of $\ccalG^c$, ${\bbTheta}_{0}^{l}, {\bbTheta}_{1}^{l} \in \mathbb{R}^{g_{l-1} \times g_{l}}$ are trainable parameters (collected in $\bbomega$), and $\sigma_l(\cdot)$ is the activation function of the $l$-th layer. 
The activation functions of the input and hidden layers are selected as leaky ReLUs, whereas a node-wise softmax activation is applied at the output layer (each row of $\bbX^L$). 
The input and output dimensions are set as $g_{0}=1$ and $g_{L}=2$.
Since $\bbcalL$ in \eqref{E:gcn} is a local operator on $\ccalG^c$, each row of $\bbX^{l}$, e.g., $\bbX_{e*}^{l}, e\in\ccalE$ can be computed through neighborhood aggregation as the following local operation on link $e$, 
\begin{equation}\label{E:gcn:local}
    \bbX_{e*}^{l} = \sigma_l \left(\bbX_{e*}^{l-1} \, \bbTheta_{0}^{l} + \left[ \bbX_{e*}^{l-1} - \sum_{u \in \mathcal{N}_{\ccalG^c}(e)}\frac{\bbX_{u*}^{l-1}}{\sqrt{d({e})d({u})}} \right]\bbTheta_{1}^{l} \right) \;,
\end{equation}
where $\bbX_{e*}^{l}\in\reals^{1\times g_{l}}$ captures the $l$th-layer features on $e$, $\mathcal{N}_{\ccalG^c}(e)$ denotes the set of (interfering) neighbors of $e$, and $d(\cdot)$ is the degree of a vertex in $\ccalG^c$.
Based on~\eqref{E:gcn:local}, the link duty cycle vector $\bbx$ can be computed in a fully distributed manner through $L$ rounds of local message exchanges between $e\in\ccalE$ and its neighbors, making our delay-enhanced BP routing a distributed algorithm.

With the estimated duty cycles of all the links, we propose two ways of setting the per-hop distance for finding the delay-aware shortest path between two nodes.
Depending on availability of the long term link rate $r_{e}, e\in\ccalE$, we define the per-hop delay distance of link $e$ as $\delta_{e}=1/x_{e}$ or $\delta_{e}=\bar{r}/(x_{e}r_{e})$, where $\bar{r}=\mathbb{E}_{e\in\ccalE,t\leq T}\left[\bbR_{e,t}\right]$ is the network-wide average link rate.
By setting the edge weights of the connectivity graph $\ccalG^{n}$ as $\bbdelta=\left[\delta_{e}|e\in\ccalE\right]$, bias $B_{i}^{(c)}$ is set as the \emph{weighted} shortest path distance between nodes $i$ and $c$ on $\ccalG^{n}$. 
This distance can be computed via distributed algorithms for weighted single source shortest path (SSSP) when a node joins the network, or weighted all pairs shortest path (APSP) in general.

\vspace{1mm}
\noindent
{\bf Complexity.}
For distributed implementation, the local communication complexity (defined as the rounds of local exchanges between a node and its neighbors) of the GNN is $\ccalO(L)$.
The distributed weighted SSSP with the Bellman-Ford algorithm \cite{bellman1958routing,ford1956network} and APSP with a very recent algorithm in~\cite{bernstein2019distributed}  both take $\ccalO(|\ccalV|)$ rounds. 
Compared to the hop distance-based methods~\cite{neely2005dynamic,georgiadis2006resource}, our approach bears additional $\ccalO(L)$ rounds of communications (and larger message size). 
Notice that $\ccalB$ can be reused over time slots until the network topology ($\ccalG^n$ or $\ccalG^c$) changes, 
which is critical for overhead reduction and scalability promotion.

\vspace{1mm}
\noindent
{\bf Throughput Optimality.}
The classical BP algorithm can stabilize the queues in the network as long as the arrival rates of flows are within the network capacity region, which is proved by Lyapunov drift theory \cite{neely2005dynamic,georgiadis2006resource}. 
The same proof also applies to BP with non-negative constant biases, as shown in \cite{neely2005dynamic,georgiadis2006resource}
and a similar proof in \cite{jiao2015virtual}.
Therefore, our approach retains the throughput optimality of the classical BP, since our proposed delay-aware shortest path bias, $B_{i}^{(c)}$, is non-negative, based on link weight $\bbdelta_e > 0$, as $\bbx_e > 0$ \forall $e\in\ccalE$ due to the softmax activation at the output layer in~\eqref{E:gcn}.

\vspace{1mm}
\noindent
{\bf Training.}
The parameters $\bbomega$ (collecting ${\bbTheta}_{0}^{l}$  and ${\bbTheta}_{1}^{l}$ across all layers $l$) of our GNN are trained on a set of routing instances defined on random network processes drawn from a target distribution $\Omega$. 
More precisely, we draw several instances (indexed by $k$) of the network topology, flows, packet arrivals, and link rates $(\ccalG^{n}(k)$, $\ccalG^{c}(k), \ccalF(k),\bbA(k),\bbR(k)) \sim \Omega$. 
For every instance, the GNN first predicts the link duty cycle vector $\bbx(k)=\Psi_{\ccalG^c(k)}(\boldsymbol{1};\bbomega)$, then biases $\ccalB(k)$ are generated by the APSP algorithm based on the link distance vector $\bbdelta(k)=\diag^{-1}(\bbx(k))\boldsymbol{1}$,
then we run the bias-based backpressure routing for $T$ time slots, and collect the schedules for each time slot, $\bbs^{k}(t)$.
We train the parameters $\bbomega$ to minimize the following mean squared error loss 
\begin{equation}\label{E:sgd_loss}
\ell(\bbomega) = \mathbb{E}_\Omega \left( |\ccalE|^{-1} \left\| \bbX^{L}(k)- \mathbb{E}_t \left(\left[\bbs^{k}(t);\boldsymbol{1}- \bbs^{k}(t)\right]\right) \right\|_{2}^{2} \right).
\end{equation}
Intuitively, by minimizing the loss in~\eqref{E:sgd_loss}, we are choosing parameters $\bbomega$ such that the softmax output of our GNN $\bbX^{L}$ is close to predicting the fraction of time that each link is scheduled.
We do not seek to learn this complex function for \emph{any specific} topology but rather we want to minimize the average error over instances drawn from $\Omega$.
In practice, we approximate the expected values in~\eqref{E:sgd_loss} with the corresponding empirical averages.
Indeed, with the collected experience tuples, we update the parameters $\bbomega$ of the GNN after each training instance, through batch training with random memory sampling, employing the Adam optimizer. 
Notice that, once trained, the GNN can be used to compute the delay-aware bias for previously unseen topologies without any retraining.
As long as the new topology, arrival rates, and link rates are similar to those observed in $\Omega$ during training, we illustrate in the next section that the trained GNN generalizes to new instances observed during testing.

\begin{figure*}[t]
\centering
 \vspace{-0.2in}
\hspace{-3.0mm}
\subfloat[]{
    \includegraphics[width=0.33\linewidth]{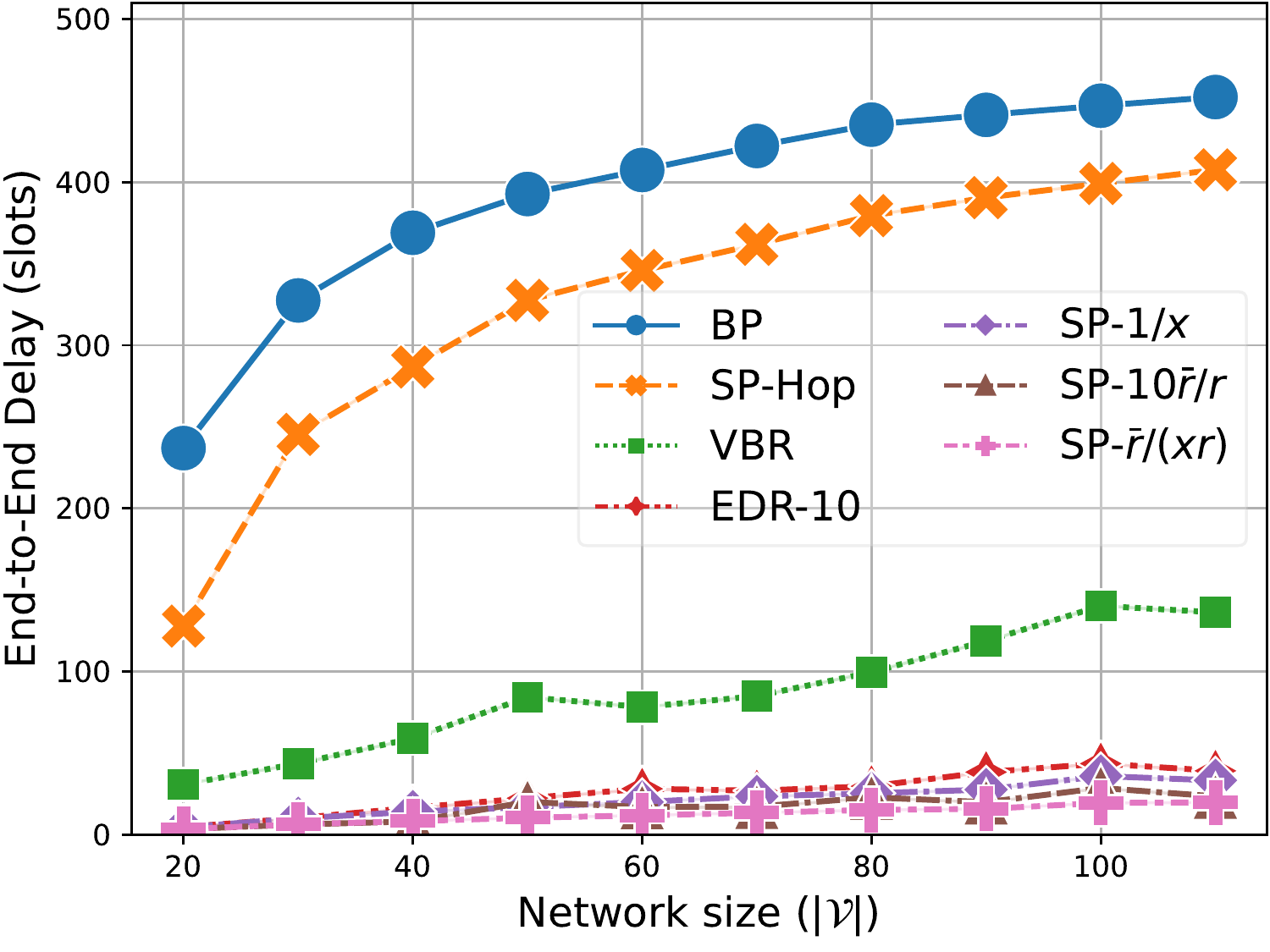}
    \label{fig:results:line}\vspace{-0.1in}
}\hspace{-2.5mm}
\subfloat[]{
    \includegraphics[width=0.33\linewidth]{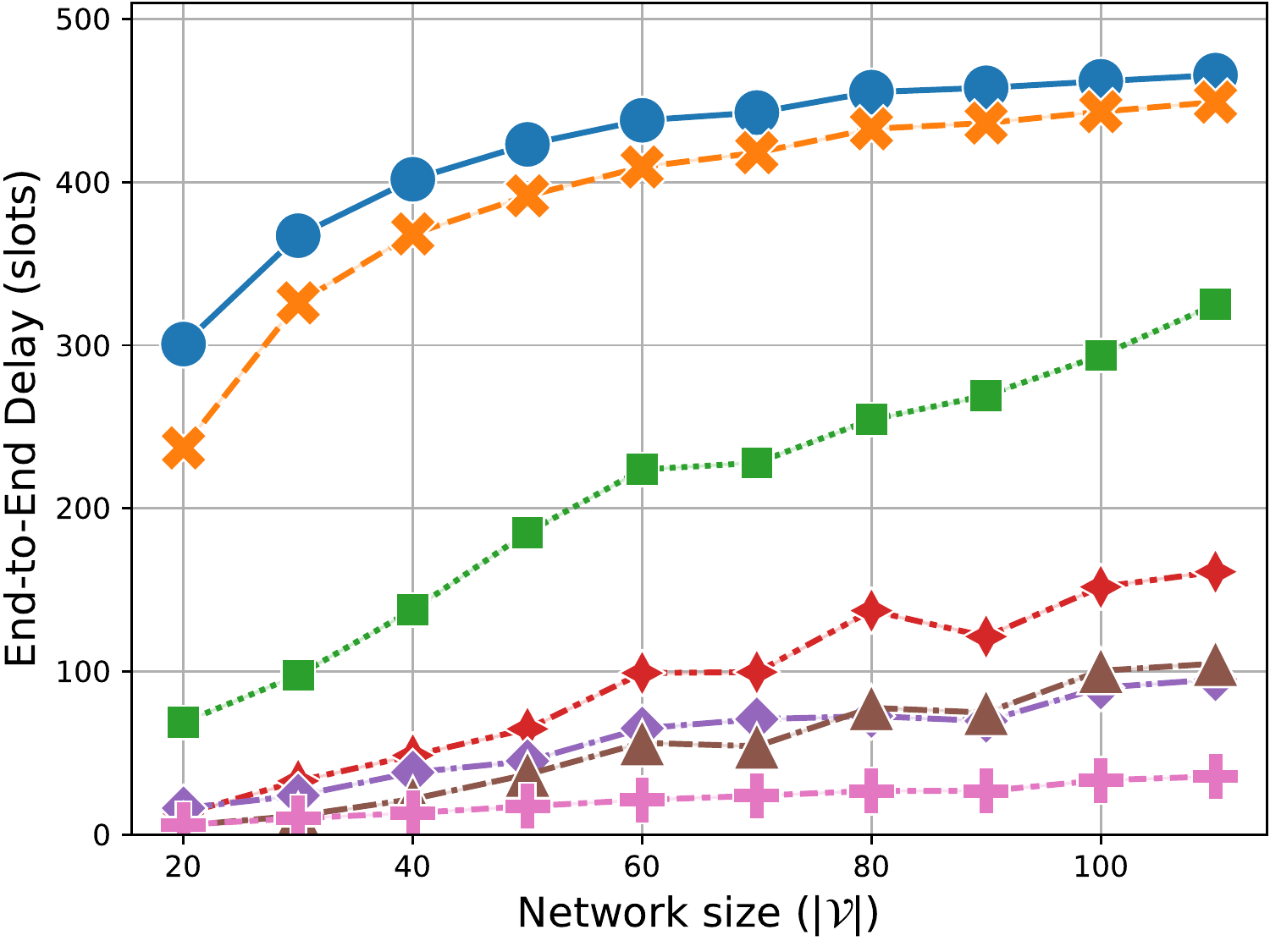}
    \label{fig:results:disk}\vspace{-0.1in}
}\hspace{-2.5mm}
\subfloat[]{
    \includegraphics[width=0.33\linewidth]{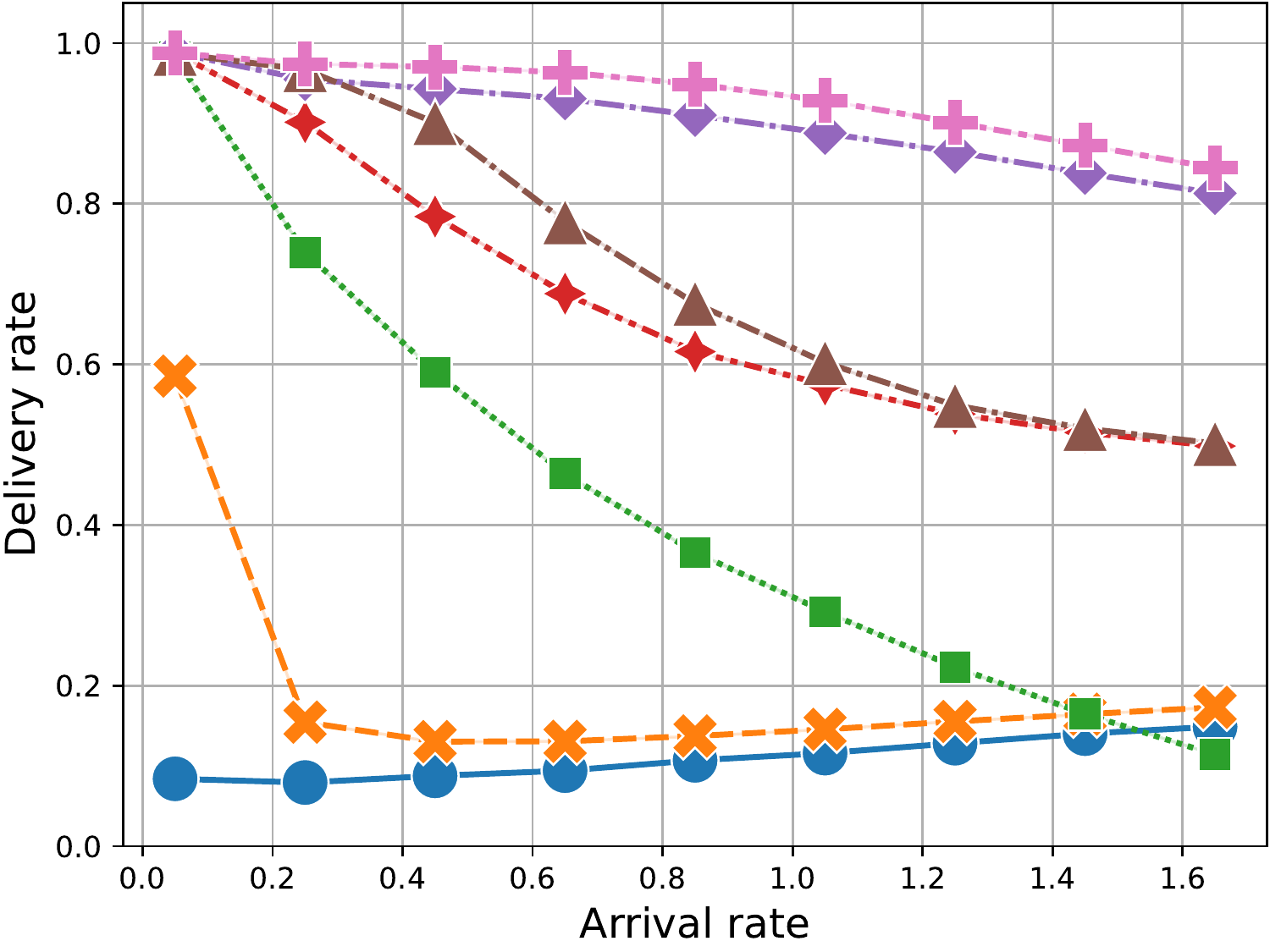}
    \label{fig:results:load}\vspace{-0.1in}
}
    \vspace{-0.15in}
    \caption{{\small Delay performance of routing schemes.
    (a)~End-to-end delay as a function of the network size under the interface conflict model.
    (b)~End-to-end delay as a function of the network size under unit-disk interference model. 
    (c)~Average delivery rate by $T=1000$ as a function of the arrival rate on random networks of $100$ nodes with unit-disk interference model.
    }
    } 
 \label{fig:results}    
 \vspace{-0.15in}
\end{figure*}

\section{Numerical experiments}
\label{sec:results}

We evaluate different biased BP routing algorithms on simulated wireless multi-hop networks. 
As explained in Section~\ref{sec:solution}, we consider two variants of our proposed delay-aware bias where the weights of each link are respectively defined as $\delta_{e}=1/x_{e}$ (denoted by SP-$1/x$) and $\delta_{e}=\bar{r}/(x_{e}r_{e})$ (denoted by SP-$\bar{r}/(xr)$). Recall that $x_e$ is the predicted link duty cycle and $r_e$ the long-term link rate. 
The competing benchmarks include the basic BP, virtual queue-based BP (VBR) \cite{jiao2015virtual}, EDR~\cite{neely2005dynamic,georgiadis2006resource} with a bias defined by the shortest hop distance, including $\delta=1$ (SP-Hop) and $\delta=10$ (EDR-10), and BP based on shortest path bias defined by link rate, $\delta=10\bar{r}/r$ (denoted by SP-$10\bar{r}/r$).
The parameters of VBR and EDR are selected according to~\cite{jiao2015virtual,cui2016enhancing}.

The simulation is configured as follows.
Every instance of a wireless multi-hop network is generated by a 2D point process with a given number of nodes $|\ccalV|\in\{20,30,\dots,110\}$ uniformly distributed in the plane with a constant density of $8/\pi$.
A simplified scenario of a single channel is simulated.
A link is established between two nodes if their distance is within $1.0$, and the two conflict models previewed in Section~\ref{sec:problem} are considered: 
for the interface conflict model the conflict graph has an average degree of $12.4$ whereas for the unit-disk interference model, the average conflict degree is $34.6$. 
Notice that all the conflicts in the former model are included in the conflicts of the latter.
Ten instances of such random networks are generated for each $|\ccalV|$.
On each randomly generated network, 10 random test instances are generated. 
Each test instance contains a number (uniformly chosen between $\lfloor 0.15 |\ccalV|\rfloor$ and $ \lceil 0.30|\ccalV|\rceil$) of random flows between different pairs of sources and destinations, where the exogenous packet arrival follows a Poisson process with a uniformly distributed arrival rate  $\lambda(f) \sim \mathbb{U}(0.2,1.0)$ for every $f\in\ccalF$. 
Each test instance also includes a realization of uniformly distributed long-term link rates, $r_{e} \sim \mathbb{U}(10,42)$, which are the expected values of the real-time link rates, $\bbR_{e,t} \sim  \mathbb{N}(r_{e}, 3)$.
The synthetic networks seek to represent wireless networks with uniformly distributed users of identical omnidirectional transmit power.
The link rates are configured to capture fading channels with lognormal shadowing.
Our proposed biases are generated based on a 5-layer convolutional GNN ($L=5$, $g_{l}=32,l\in\{1,\dots,4\}$), trained on a set of 100 random networks with $|\ccalV|\in\{20,30,\dots,60\}$, and the flows and link rates are configured similarly to the test settings.\footnote{Training takes 5 hours on a workstation with a specification of 16GB memory, 8 cores, and Geforce GTX 1070 GPU. The source code is published at \url{https://github.com/zhongyuanzhao/dutyBP} 
}

Our simulations lasts for $T=1000$ time slots, where we compare the average end-to-end delay of packets attained by the 7 routing schemes considered.
{To be conservative, we treat the delay of an undelivered packet as $T-t_0$, where $t_0$ is the time it arrived at the source node.}
The estimated end-to-end delay of routing schemes as a function of the network size for both conflict models are presented in Figs.~\ref{fig:results:line} and~\ref{fig:results:disk}.
The delays increase with network size for all methods due to the longer average hop distances of flows on larger topologies.
The vanilla BP performs the worst due to its low delivery rate (fraction of packets delivered by the end of the simulation) that ranges from $0.56$ in small networks ($|\ccalV|=20$) to $0.11$ in larger networks ($|\ccalV|=110$). 
By introducing hop distance as bias, SP-Hop can improve the delay of BP, and by scaling this bias by a factor of 10, EDR-10 can significantly reduce the delay of BP.
The SP-$10\bar{r}/r$ method can further reduce delays of EDR-10 by $1/3\sim1/2$ through additional information of the long-term link rates across the network.
Leveraging our GNN predictions of the scheduling duty cycles of links, SP-$1/x$ and SP-$\bar{r}/(xr)$ can further reduce the delays over EDR-10 and SP-$10\bar{r}/r$, respectively, where the improvements are more visible for the scenarios with higher average conflict degrees; see Fig~\ref{fig:results:disk}.
Contrary to the results in~\cite{jiao2015virtual}, the delay of VBR is much higher than EDR-10 in our tests, mainly due to the fact that VBR is designed for wireless sensor networks with only one sink, whereas flows in our simulation have different destinations.
The test results show that our approach can generalize to networks larger than those seen during training, and can improve the delay of BP routing by offloading traffic from hot-spots congested by many interfering neighbors.
Notice that shortest path biases (in hops) are myopic to this kind of information.
Moreover, our GNN can learn the scale of the needed bias given the interference density of the network, alleviating the need of careful parameter tuning needed in the scaling of other methods, such as EDR-10 and SP-$10\bar{r}/r$.

Next, we evaluate the delay performance of routing schemes under different traffic loads, on 10 random networks of 100 nodes, under the unit-disk interference model.
For each random network, 10 realizations of random flows and random link rates are generated like in the previous experiment, except that all the flows have identical arrival rate $\lambda \in \{0.05,0.25,\dots,1.65\}$.  
The delivery rates of different routing schemes after $T=1000$ time slots are presented in Fig.~\ref{fig:results:load}.
Our approaches achieve the highest delivery rates under heavier network loads. 
This shows that we can improve both the delay and the delivery rate (number of packets delivered per unit of time) over existing biased BP algorithms across different traffic loads by leveraging our learning-based framework.

\section{Conclusions}
\label{sec:conclusions}
We improve the shortest path bias-based BP routing by replacing the well-established hop distance with a delay-aware distance.
This delay-aware distance is computed using our predictions of the scheduling duty cycle of links given by a GNN. 
This solution inherits the simplicity, low per-slot overhead, and throughput optimality of BP routing with shortest path bias. 
Experiments show that our approach can outperform other BP algorithms based on pre-defined biases in terms of end-to-end delay and delivery rate across different interference densities and traffic loads, and exhibits good generalizability to larger networks. 
\vfill\pagebreak



\bibliographystyle{ieeetr}
\bibliography{strings,refs}

\end{document}